# Business Process Management: Saving the Planet?

## Shahrzad Roohy Gohar
UQ Business School
The University of Queensland
St Lucia, Queensland
Email: sh.roohygohar@business.uq.edu.au

## Marta Indulska
UQ Business School
The University of Queensland
St Lucia, Queensland
Email: m.indulska@business.uq.edu.au

## Abstract

Organisational and government concerns about environmental sustainability (ES) are on the increase. While a significant amount of research from a wide range of domains has addressed various ES challenges, intuitively, Business Process Management (BPM), with its focus on process improvement and process performance measurement, has much to offer the ES field. In this paper we aim to understand the BPM research contribution to ES, including specific Environmental Performance Indicators (EPI), and the BPM concepts that have been utilized in the ES context. To this end we conduct a systematic literature review to capture prior research focused on BPM and ES, coding the articles according to their contribution to EPIs and other ES concepts, while also contrasting their focus with main challenges identified in industry reports. Our study identifies which EPIs have been addressed in prior BPM research and highlights areas of future contribution.

## 1 Introduction

Sustainability is a global issue that requires a clear and bold response from organizations and governments. It is a broad and complex topic that spans environmental, social and economic issues. Sustainable development, introduced in 1987 by the World Commission for Environment and Development (WCED) as "Our Common Future" (Brundtland 1987), is development that meets the needs of the present without compromising the ability of future generations to meet their own needs. Initially, sustainability was defined by the rate at which renewable energy resources could be extracted without causing damage to the ecosystem (Dincer 2000). Today's sustainability agenda, however, has expanded to consider three aspects, *viz.* economic, social and environmental (Goodland 1995). Economic sustainability seeks to maintain economic and man-made capital (Goodland 1995). Social sustainability refers to equity and equal redistribution of resources (1995). Environmental Sustainability (ES) provides insight and solutions for sustaining natural capital (Goodland 1995). These three aspects of sustainability, *viz.* environmental, social and economic, are known as the "triple bottom line" (Elkington 1997; Savitz 2012) and are the three pillars of corporate sustainability (Elkington 1997; Epstein and Roy 2001; Savitz 2012).

Focussing on ES, it is estimated that information and communication technologies could deliver up to a 15% reduction of carbon emissions by business by 2020 (Lacy et al. 2010), thus having a significant contribution to ES. Information Systems (IS) and Business Process Management (BPM) are both well placed to offer solutions for ES. BPM specifically, with its focus on improvement and process monitoring, is relevant in the ES context. Thus, while the academic community has explored ES from a variety of angles, our interest lies in the potential contribution of BPM research. BPM is a management approach that focuses on improving organizational efficiency and effectiveness (Hammer 2010) and optimizes technology use and employee performance (Vom Brocke and Rosemann 2010). One of its main aims is to optimize the use of organizational resources e.g. time, raw materials, or more specific benchmarks for organizational performance (Weske 2012). Such resources become Key Performance Indicators (KPI) and can be measured at the process and organizational levels through BPM (Parmenter 2010; Wetzstein et al. 2008). Environmental Sustainability is measured using Environmental Performance Indicators (EPI). The main EPIs commonly captured are those relating to air pollution, water quality and consumption, energy and fuel consumption, material recycling and waste management (Hammond 1995; Yale Center for Environmental Law & Policy 2014; Young and Rikhardsson 1996). Intuitively, there is a strong connection between BPM and ES due to BPM's process-oriented approach and facilitation of measuring EPIs at the process level (DeToro and McCabe 1997; Ghose *et al.* 2010). Indeed, authors have previously explored this relationship. The term 'Green BPM' used by Seidel et al. (2012) relates to understanding, documenting, executing and constantly improving processes with a focus on their environmental impacts; hence Green BPM supports designing and implementing Environmentally Sustainable processes. Opitz et al. (2014a) defines Green BPM as sum of all IS-supported management activities contributing to reduction of environmental impact of business processes, including design, improvement, process lifecycle and operational steps. Accordingly we adopt the same terminology and refer to BPM within an ES context as 'Green BPM' and are motivated to explore to what extent Green BPM has been studied and how it has contributed to the ES body of knowledge. Thus, we conduct a literature review aimed at exploring the extent to which the BPM research community has contributed to ES. In doing so, we present the first study to analyse and classify BPM research contributions based on EPIs and organizational factors relating to ES (Brooks et al. 2012). In addition, we are interested to explore what aspects of BPM have been applied in these studies and how these studies have addressed industry needs. Thus, our study intends to answer the following questions:

i. How has BPM research contributed to ES?

ii. What BPM concepts have been applied in ES research?

iii. What BPM related research gaps are evident based on ES industry needs?

In answering the above questions we aim to provide a single point for new Green BPM researchers to understand current contributions to EPIs and ES organisational factors. In addition, we identify gaps in the research that are yet to be addressed despite being of relevance to industry.

The remainder of the paper is organized as follows. Section 2 discuses related studies that aimed to identify BPM contributions to ES. Section 3 summarizes our study design and methodology. Section 4 explores the research questions and summarizes BPM's contribution to ES thus far, also identifying topics for future research. The paper concludes in section 5 with a summary of all findings and future research directions.

## 2　Related Work

A systematic literature review is essential to capture the state of the art prior to a theoretical contribution (Webster and Watson 2002). In the following we discuss relevant prior literature reviews and differentiate our research from prior published works.

Several authors over the last decade have attempted to explore Green BPM research contributions. Stolze et al. (2012) conducted a review of 2006-2011 literature from English and non-English sources in the Information Systems discipline. They classified identified research publications based on high-level keywords (e.g. 'Green IT', 'Green IS', 'Sustainable', 'Business Process' etc). While the study provides a high level of classification, it does not explore in depth the BPM concepts typically seen in ES research, nor the contribution of BPM to various EPIs.

Opitz et al. (2014b) explored the acceptance of ES in BPM literature and studied the potential of developing models and methods to measure an organization's ability to implement Green BPM. Their study identified 127 papers of which 11 were relevant to BPM. The relevant papers were classified based on 'attitude', 'strategy', 'governance', 'modelling', 'optimizing', 'monitoring' factors and further proposed a conceptual Green BPM readiness model motivated by the Green ICT Readiness Model (Wabwoba et al. 2013). Nevertheless, the study did not explore EPIs to which contributions were made and had a more narrow focus. Opitz et al. (2014a) then extended the review with a study of 56 articles searching for definitions and concepts related to Green BPM. This follow-up study classified literature based on categories of 'Green IT/IS', 'BPM', 'Green BPM' and classified the papers based on their contribution to subcategories of 'Reduce Environmental Impact', 'Monitoring', 'Economical', 'Cultural Change' and 'Definition'. However, the extent of theoretical contributions, detail of BPM and ES concepts covered in the study remains unspecified, and a focus on EPIs is lacking.

There is a common strategy in the above studies – one of gap identification in literature. The authors focused on articles from prominent IS and BPM journals and highly ranked IS and BPM conferences, with one study seeking further sources in non-English international conferences and journals. Identification and analysis of BPM research contributions were limited to the abovementioned concepts. Our study extends these efforts to provide deeper and more specific insights into the extension of BPM concepts towards ES. We identify industry relevant ES concepts, such as EPIs and organizational factors relating to ES (Dada et al. 2013) and analyse to what extent BPM research has contributed to these concepts. Through this analysis and related discussion we offer deeper insights into research on Green BPM and discuss to what extent the research addresses the concerns cited in authoritative ES industry reports.

# 3　Study Design

## 3.1　Scoping

Although IS and BPM are comparatively old disciplines, ES is a relatively new topic in these domains. The earliest ES related academic publication in IS was published in 2006 and there is hardly any published academic research focusing on ES in BPM prior to 2005. Accordingly, our scope of study covers the period of time from 2005 to date. In an effort to identify as many relevant articles as possible, we relied on Google Scholar rather than purely IS or BPM academic publication outlets, although we used such a data set initially to identify relevant keywords. In addition to identifying academic papers, we used Google to identify authoritative industry reports on ES issues and challenges. The search strategies for both types of contributions are outlined in the following sub-section.

## 3.2　Search Strategy

To identify search terms relevant for our study we first explored ES related keywords, variations and word-stems in use specifically by the IS and BPM community. To this end we used a set of publications in prominent IS journals and high quality conferences, since most IS and BPM publications with or without ES focus are published in these outlets. Our data set includes the Association for Information Systems (AIS) senior scholars basket of 8 journals[1], the Business Process Management Journal (BPMJ) and top IS and BPM community conferences such as PACIS, ECIS, ICIS, AMCIS, ACIS and BPM since 2005.

Numerous publications were studied for frequently used ES related keywords in IS and BPM research. We identified 'Green', 'Environmentally Sustainable', 'Sustainability' and 'Sustainable', which are quite broad in nature. We also considered stemmed searches using derivations and variations of the terms 'Environmental' and 'Sustainability'. Through keeping the search terms broad and combining them with Boolean operators as shown in Table 1, we identified a pool of papers to further analyse so we could identify more specific ES related keywords used by IS and BPM researchers. We observed through a manual review that some papers did not specifically mention ES in their title, yet still focused on ES (e.g. such as 'carbon-footprint ($CO_2$)', Energy, 'GHG emission' and Ecological). Therefore, we included these terms to ensure inclusiveness of our later search using Google Scholar. Terms such as 'environmentally-aware', 'carbon-aware', 'energy-aware' were also identified in the initial identification of search terms. Through this process of identification and collection of most used ES terms in IS and BPM publications, we derived a list of relevant search terms (see Table 1).

| **Broad BPM/IS search terms** | **Boolean Operators** | **Broad ES terms** |
| --- | --- | --- |
| Information Systems (IS) | AND | Environmentally Sustainable, |
| Business Process Management (BPM) | | Environmental Sustainability, |
| Business Process Reengineering (BPR) | | Green |
| | | Environmental, Envairomentally, Environment*, Sustainable, Sustainability, Sustain* |
| **Narrow BPM/IS search terms** | | **Narrow ES Search terms** |
| Process | | Carbon-footprint, GHG Emission, Energy, Ecological, environmentally-aware, carbon-aware, energy-aware |

*Table 1. Search Terms and Operators*

---

[1] European Journal of Information Systems (EJIS), Information Systems Journal (ISJ), Information Systems Research (ISR), Journal of AIS (JAIS), Journal of Information Technology (JIT), Journal of Management Information Systems (JMIS), Journal of Strategic Information Systems (JSIS), MIS Quarterly,

Having identified a set of search keywords, we conducted the combinations of stemmed keyword searches using Google Scholar and collected relevant publications into a primary collection of literature. Our searches were initially conducted on article titles only as we considered that articles with titles containing combinations of our BPM and ES keywords would have a central focus on ES and BPM. We then performed iterative forward and backward searches, exploring bibliographic references and authors in retrieved publications.

Our search identified 260 journals and conference papers spanning Green BPM, Green IT and Green IS. By maintaining the same approach of looking into article titles at first, we categorized the findings into three main tiers. Thirty-six publications matching ES relevant keywords AND Business Process/BPM are considered our tier 1 publications. Subsequently, 112 papers with titles containing ES related terms AND IS are tier 2 and 97 papers focused on ES AND IT are categorized as tier 3. We excluded tier 2 and 3 in order to keep the focus of the study on ES AND Process/BPM only. The inclusion and exclusion criteria and paper tiers are presented in Table 2.

| Inclusion Tier and Criteria |
| --- |
| Tier 1: Green BPM – articles with focus on BPM and ES (36 papers) |
| Industry reports with a focus on ES (6 reports) |
| **Exclusion Tier and Criteria** |
| Tier 2: Environmentally Sustainable IS (112 papers) |
| Tier 3: Environmentally Sustainable IT (97 papers) |
| Articles did not match the inclusion criteria |
| Non-English articles |
| Articles containing search terms such as BPM AND Process but not relevant to our focus and inclusion criteria |

*Table 2. Inclusions and Exclusions*

To ensure an exhaustive collection of BPM focused articles, and assuming potential IS focused articles might contain BPM related concepts, we conducted a full-text search for the term "Process" in the 112 tier 2 articles, which resulted in 20 matching articles. However, a manual review revealed none had BPM as their core focus; therefore they were not included in our analysis. Accordingly, our analysis focuses on 36 highly relevant papers identified with Google Scholar and resulting forward and backward searches.

To identify global and industrial ES challenges and to understand where the ES focus lies in industry, we performed a search in Google using keywords of 'Environmentally Sustainable', 'Environmental Sustainability' and 'Environmental Performance Indicators (EPI)'. We identified and incorporated 6 major authoritative published agendas and annual reports regarding ES in industry within the timeframe of 2005-2015, including reports from global bodies (e.g. United Nations (UN)), main industry bodies, as well as the Australian government for a local indication (see section 4), as follows:

- Australian Government's strategies and actions to introduce low carbon emission initiatives and improve the sustainability of ICT operations (Australian Government Sustainability Plan 2010-2015 (Commonwealth of Australia 2010)
- Accenture's CEO study and reflections on challenges and the impact of the journey toward a sustainable economy in: A New Era of Sustainability - UN Global Compact by Accenture CEO Study 2010 (Lacy et al. 2010)
- 2014 Annual report by United Nations on the Environment Programme highlights global development significance and remaining challenges ahead to overcome climate change and remaining sustainability issues (United Nations Environment Programme 2014).
- The Synthesis Report on Climate Change 2014 presents findings of three working groups that contributed to the Fifth Assessment Report (AR5) of the Intergovernmental Panel on Climate Change (IPCC), which is by far the most complete assessment of climate change undertaken by the IPCC (IPCC 2014).
- The Sustainable Australia Report 2013, which demonstrates emerging issues and major trends for Australia's sustainability, including Environmental Sustainability (National Sustainability Council 2013).
- Environmental Sustainability and Industry - Road to a Sustainable Future is the largest survey of environmental practices developed from findings of National Sustainability Council 2013 conducted by the Australian Industry Group in conjunction with Sustainability Victoria (Australian Industry Group 2007).

### 3.3 Coding and content analysis

Subsequent to identifying and collecting the relevant literature, we coded the collection of 36 papers to uncover dominant concepts and ideas discussed in literature. We identified concepts, their significance, frequency and existence across selected literature. We looked for search terms, relevant phrases, theoretical constructs and proposed research artefacts. Specifically, we looked for relevant BPM concepts addressed, advanced and developed as well as any of ES concepts addressed, studied, tested and/or implemented using process-oriented methods. To do so, we first identified a set of coding criteria, implemented in an Excel spreadsheet.

First, we began with basic codes, resulting in spreadsheet columns capturing "Title", "Year of Publication", "Main Contribution", "BPM Concept", "Research Methodology", "Data Collection Method", "Data Analysis Method", "Assumptions" and "Limitations".

Our ES and BPM specific coding criteria were based primarily on sustainability concepts identified from key papers in the Sustainability domain and frequently mentioned concepts in BPM academic publications (Australian Industry Group 2007; Epstein and Roy 2001; Goodland 1995; Hammond 1995; Hoesch-Klohe and Ghose 2012; Jasch 2000). We recorded main themes of papers based on "Type of Sustainability (Economic, Social, Environmental)", EPIs such as "Energy Consumption", "CO2 footprint", "GHG Emission", "Waste management", "Water consumption" and "Recycling". Similarly organizational factors related to ES were identified and recorded as "Management, "Strategy" and "Culture".  For organisational factors, we classified studies on management of structure, practices, operations, inter-organizational collaborations to support ES as "Management" factors; decision-making, internal and external policies as "Strategy" factors; and ES organizational culture as "Culture" factors in our study (Dada et al. 2013; Sharma 2000; Wesumperuma et al. 2011).

All 36 publications were iteratively examined for presence of the abovementioned concepts. While the EPIs and organisational factors were identified from literature prior to coding the 36 articles, the BPM concepts emerged iteratively through consolidating the "BPM Concept" code and recoding all of the papers iteratively as these concepts emerged. The final set of BPM related concepts are as follows:

 "BPM Lifecycle Extension": relates to suggested models and additional components to current BPM lifecycle models (Rosemann and vom Brocke 2010; Van Der Aalst 2004; Weske 2012) aiming to design, analyse, model and validate business processes based on ES objectives i.e green BPM Lifecycle (Nowak et al. 2011c; Opitz et al. 2014b; Recker et al. 2012)

"BPM Architecture Extension": refers to proposed models and tools to improve business processes, operations and management architecture of organizations based on targeted environmental sustainability goals (Harmon 2010; Nowak et al. 2011c; Nowak et al. 2011d).

"Capability Maturity Model Extension": refers to extended theoretical dimensions for organizations to measure their current maturity level based on ES/sustainability dimensions in order to improve in those dimensions (Cleven et al. 2012; Seidel et al. 2012).

"Process Performance Measurement Method Extension": refers to developed and/or adopted performance measurement methods to capture environmental process performance based on EPIs (Ardagna et al. 2008; Hoesch-Klohe and Ghose 2010; Nowak et al. 2011a; Recker et al. 2012; Thies et al. 2012; Wesumperuma et al. 2013).

"Process Modeling Extension": refers to additional modelling elements in process modelling notations to enrich process models with ES concepts (Hoesch-Klohe and Ghose 2012; Hoesch-Klohe et al. 2010; Recker et al. 2012; Wesumperuma et al. 2011).

"Process Reengineering" and "Process Design": refer to techniques and methods to assist organizations in designing and reengineering (Hammer and Champy 1993) their business processes based on environmental objectives, e.g. reducing CO2 or GHG emission (Ghose et al. 2010; Hoesch-Klohe and Ghose 2010; Nowak et al. 2011a; Nowak et al. 2011d).

"Process Optimization": (re-design, improve and adopt) refers to developed concepts based on iterative stages of monitoring, redesigning and improvement of business processes towards ES organizational objectives (Ghose et al. 2010; Houy et al. 2012; Nowak and Leymann 2013; Wesumperuma et al. 2011).

"Definition of Green BPM Extension" includes attempts to clearly define Green BPM based on past definitions of BPM (Weske et al. 2004) and proposed environmental performance aspects and environmental indicators of processes (Ghose et al. 2010; Opitz et al. 2014a; Seidel and Recker 2012).

## 4　Results

Of the 36 academic publications, 21 were conference articles, 4 were journal articles and 11 were book chapters published between 2005-2015 (see Figure 1). There is hardly any research available on BPM in the ES context prior to 2008. The results indicate some initial interest in ES from the BPM community but there is no evident trend.

Following multiple iterations reading the papers, we found that the main contributions from BPM in the ES context were through process optimization, process performance measurement methods, and process design/reengineering (see Figure 2). For example, classified as "Process Optimization", Pernici et al. (2008) present a "context-aware" and "purifier-based" approach to ES, aiming to optimize energy consumption without compromising efficiency. Ghose et al. (2010) suggest a theoretical roadmap to optimise processes through the Carbon Modelling Framework. Nowak et al. (2011d) propose and test an architecture, and a four-phase approach to process optimization by defining and measuring ecological characteristics, identifying their environmental impact, and helping them to develop appropriate adaptation strategies to optimize their environmental impact while keeping organizational effectiveness.

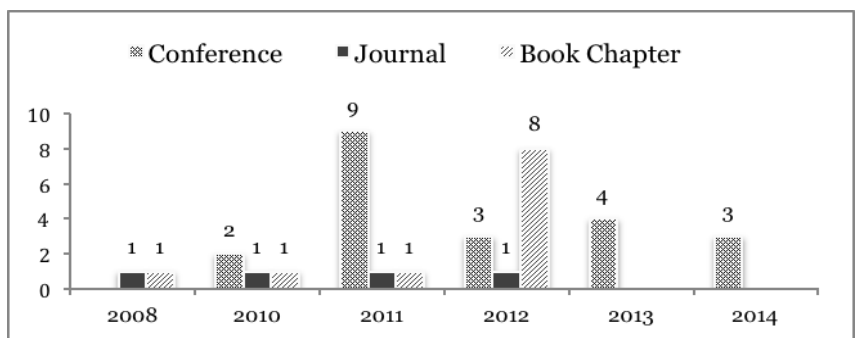

*Figure 1. Distribution of publications per type per year*

Research extending or applying "Process Performance Measurement Methods" has typically focused on EPIs of "CO2 footprint", "Energy Consumption" and "GHG emission" (Ardagna et al. 2008; Hoesch-Klohe and Ghose 2010; Hoesch-Klohe et al. 2010; Recker et al. 2012; Thies et al. 2012).

In terms of process design, Nowak et al. (2011b) identified frequently used patterns by organizations to improve sustainability. The study also focused on organizational factors of ES, such as management. The study was later extended (2013) to help organisational stakeholders identify patterns and design environmentally-aware business processes.

Seidel et al. (2011) identified the need to extend the BPM lifecycle into a 'Green BPM lifecycle'. Twenty-two percent of the relevant papers have contributed in this manner.

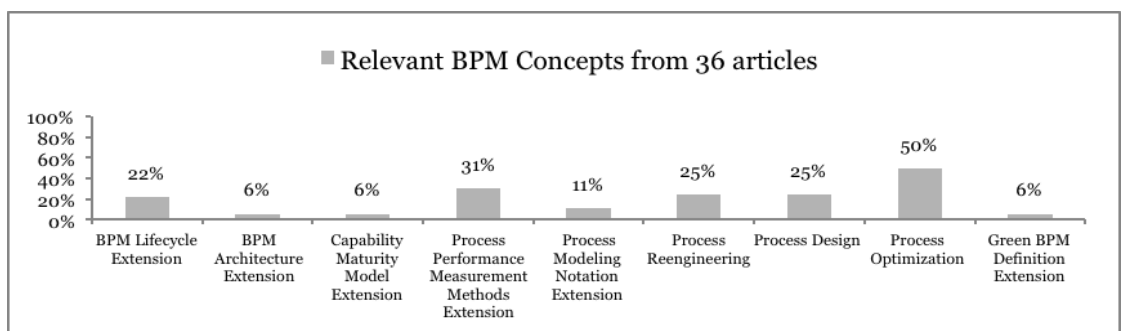

*Figure 2. Relevant BPM Concepts*

The main ES focus of these 36 papers was on several organisational factors, specifically management and strategy factors (see Figure 3). Research exploring Green BPM from an organizational perspective has mostly focused on "Management" and "Strategy" factors. Seidel et al. (2010) identify management collaboration, strategy and bottom-up organizational support to play major role in companies managing their sustainability objectives. Nowak et al. (2011d) presents a four-layered architecture and four-phased methodology that enables organizations to define and measure their ecological characteristics, identify and visualize their environmental impact and develop adaptation strategies to optimize their environmental performance. Ahmed and Sundaram (2011) suggest a theoretical roadmap to Sustainable Business Transformation (SBT), presenting steps to strategize, plan, transform and monitor change in business environment towards sustainability. Houy et al. (2012) applies scenarios in their study on corporate sustainability by introducing Green Service Level Agreement (gSLA) in management as well as a fragment replacement method for process optimization. Cleven et al. (2012) suggests a capability maturity model for organizations to determine their current performance maturity and identify required improvements in their business processes towards sustainability. Nowak et al. (2012) uses best practices of cloud application to introduce ES adaptation strategies defined as green business process patterns. The study proposes a pattern-based approach to process adaptation and implies changes need to be extended to the application and infrastructure layer.

In terms of the core EPIs, most papers were focussed on Energy consumption, measuring $CO_2$ footprint and GHG Emissions. This is perhaps not surprising given how much attention energy consumption and carbon footprint have received in the media over the last decade.

Focusing on energy consumption performance indicators, BPM related studies have proposed energy-aware process-based applications (Ardagna et al. 2008), a context-aware purifier-based approach to ES (Pernici et al. 2008), examining KPIs, complex event processing and case-management application to energy management systems (Watson et al. 2012) and an integration model for energy consumption of IT components and business processes (Reiter et al. 2014). In terms of $CO_2$ and GHG emission related research, contributions include an application of the Abnoba algebraic framework to multiple heterogeneous dimensions for process improvement (Hoesch-Klohe and Ghose 2010), a network-centric solution for sharing and using EPIs (Thies et al. 2012), and measuring and monitoring suitable metrics to optimize processes while maintaining energy efficiency and good performance (Cappiello et al. 2013). Furthermore, studies by Recker et al. (2011) propose Activity-Based Emission (ABE) analysis for measuring $CO_2$ of processes and an approach for documenting the carbon footprint of business processes in an extended business process model (Recker et al. 2012).

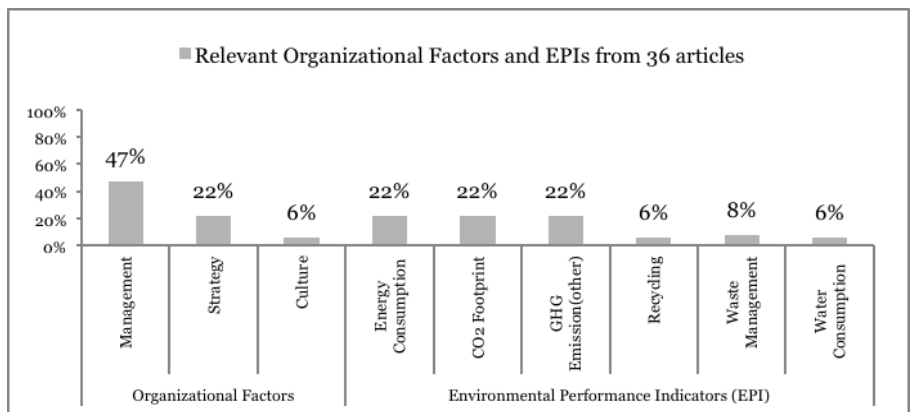

*Figure 3. Relevant Organizational Factors and EPIs*

In addition to the analysis of 36 BPM academic articles, 6 reports were collected and analysed from global institutions such as United Nations (UN), large industry bodies and Australian Government. Analysis of these reports indicates that they consider the majority of the core EPIs (see Table 3). This finding implies a global interest in managing specific EPIs.

| Reports | Organizational Factors | | | Environmental Performance Indicators (EPI) | | | | | |
|---|---|---|---|---|---|---|---|---|---|
| | Management | Strategy | Culture | Energy Consumption | CO2 Footprint | GHG Emission (Other) | Recycling | Waste Management | Water Consumption |
| (Commonwealth of Australia 2010) | 1 | 1 | – | 1 | 1 | 1 | 1 | 1 | 1 |
| (Lacy et al. 2010) | 1 | 1 | 1 | 1 | 1 | 1 | 1 | 1 | 1 |
| (United Nations Environment Programme 2014) | 1 | 1 | 1 | 1 | 1 | 1 | – | 1 | 1 |
| (IPCC 2014) | 1 | 1 | 1 | 1 | 1 | 1 | 1 | 1 | 1 |
| (National Sustainability Council 2013) | – | 1 | 1 | 1 | 1 | 1 | 1 | 1 | 1 |
| (Australian Industry Group 2007) | 1 | 1 | 1 | 1 | 1 | 1 | 1 | 1 | 1 |
| Total | 5 | 6 | 5 | 6 | 6 | 6 | 5 | 6 | 6 |

*Table 3. Focus of Industry reports per ES Organizational Factors and EPIs*

Our aim in this research is also to uncover to what extent BPM literature has contributed to these commonly cited EPIs. In the following subsections we discuss the contribution to each ES concept, while Table 4 provides a summary of the number of articles that contribute to a particular ES concept through a specific BPM concept.

| BPM Concepts | Organizational Factors | | | Environmental Performance Indicators (EPI) | | | | | |
|---|---|---|---|---|---|---|---|---|---|
| | Management | Strategy | Culture | Energy Consumption | CO2 footprint | GHG Emission (Other) | Recycling | Waste Management | Water Consumption |
| BPM Lifecycle Extension | 2 | 2 | - | - | 3 | 1 | - | 1 | - |
| BPM Architecture Extension | 1 | 1 | - | - | - | - | - | - | - |
| Capability Maturity Model Extension | 2 | 1 | - | - | - | - | - | - | - |
| Process Performance Measurement Extension | 4 | 2 | 1 | 3 | 5 | 4 | 2 | 2 | 2 |
| Process Modeling Extension | 1 | - | - | - | 3 | 2 | - | - | - |
| Business Process Reengineering | 4 | 2 | - | 3 | 1 | 2 | 1 | 1 | 1 |
| Process Design | 5 | 1 | 1 | 4 | 5 | 4 | 2 | 2 | 2 |
| Process Optimization | 8 | 3 | - | 5 | 7 | 7 | 1 | 1 | 1 |
| Green BPM Definition Extension | 1 | - | - | - | - | - | - | - | - |

*Table 4. BPM concepts contributing to ES Organizational factors and EPIs*

## 4.1 Organizational Factors

Organizational factors in our study are considered as management of structure, practices, operations, decision-making, policies, inter-organizational collaboration and architecture of organizations within the context of ES (Wesumperuma et al. 2011). Organizational strategy and culture are also considered to be major factors influencing an organization's environmental performance (Dada et al. 2013; Epstein and Roy 2001; González-Benito and González-Benito 2006; Sharma 2000). All three types of factors are essential in achieving ES. New values need to be implemented across organizations at all levels, novel strategies need to be monitored and regulated by national and international bodies and planning for cultural change is a requirement to achieve targeted environmental performance (Lacy et al. 2010).

From 36 BPM research publications in our scope of study, we found 17 articles acknowledging the significance of organizational factors by using BPM concepts in the development of conceptual frameworks and models that included organizational perspectives. Table 4 summarises the application of various BPM concepts to ES related organisational factors.

## 4.2 Environmental Performance Indicators (EPI)

Energy consumption, water consumption, CO2 footprint, Greenhouse Gas (GHG) Emission, Waste management and Recycling of materials are the six core Environmental Performance Indicators measured by organizations. CO2 has been identified to play a major role in climate change and global warming; therefore it has received much attention in environmental frameworks across the globe (Goodland 1995; Hammond 1995; Jasch 2000), however there is a need to focus on all EPIs.

**Energy Consumption** is a main agenda cited in all of our 6 identified industry/government reports, including the United Nations Environment Programme (2014) as well as Australian ICT Sustainability plan by Commonwealth of Australia (2010). It requires an environmental management strategy to audit and control energy consumption. From the 36 identified research articles, 8 (22%) contain contributions to energy consumption management. These contributions are conceptual in nature and include a focus on energy-aware applications for minimizing energy consumption (Ardagna et al. 2008), purifier-based approaches (Pernici et al. 2008), network-centric solutions (Thies et al. 2012), application of Abnoba Algebraic framework to multiple heterogeneous dimensions (Hoesch-Klohe and Ghose 2010), and process improvement methods to support measuring and monitoring performance and energy efficiency and a conceptual integration model for energy consumption from IT components to business processes (Reiter et al. 2014). Further, Houy et al. (2011) emphasizes reduction of energy consumption in processes by applying techniques and solutions from BPM.

**CO2 and Greenhouse Gas Emission** are thought to directly influence climate change (IPCC 2014; Young and Rikhardsson 1996). CO2 emission from fossil fuels and industrial processes contributed approx. 78% of the total GHG accumulation from 1970 to 2010 (IPCC 2014)**.** Continued emission of gases will increase the possibility of pervasive and severe impacts for the population and the ecosystem by increasing temperatures of the earth's surface and oceans (IPCC 2014). The United Nations Environment Programme (2014) has visioned a roadmap for emissions cuts and has estimated that climate change adaptation costs will reach $300 billion per year by 2050. Climate change concern about GHG emissions also continues to grow (Lacy et al. 2010). From the 36 identified BPM research articles, 8 articles (22%) contribute to measurement and management of CO2 and GHG emissions. These contributions include a conceptual framework for carbon-aware process improvement (Hoesch-Klohe and Ghose 2010), a conceptual roadmap to optimise the carbon modelling framework (Ghose et al. 2010), activity-based emission analysis for measuring CO2 in processes (including modelling notation extension, specifically BPMN) (Recker et al. 2011; Recker et al. 2012), network centric solutions (Thies et al. 2012), conceptual advancement of methods for measuring and monitoring process performance based on EPIs such as CO2 (Cappiello et al. 2013) and theoretical principles for capturing, measuring, modelling and reporting GHG emissions (Wesumperuma et al. 2013).

**Recycling, Waste Management and Water Consumption** have received less focus from the BPM research community. Out of 36 publications, two focus on waste management; with recycling and water consumption management having one related article each. Hoesch-Klohe and Ghose (2010) discuss the potential of considering extended EPIs in the application of the Abnoba Algebraic framework for process optimization. Thies et al. (2012) propose a network-centric solution to share and use of EPIs in business processes. Houy et al. (2011) emphasize the opportunity of waste reduction in processes by applying techniques and solutions from BPM. Thus, while the main global and industry reports indicate the significance of these three EPIs in achieving ES (Australian and New Zealand Environment and Conservation Council State of the Environment Reporting Task Force 2000; Australian Industry Group 2007; Commonwealth of Australia 2010; IPCC 2014; United Nations Environment Programme 2014), a substantial response from the BPM research community is yet to come.

## 4.3 Research gaps vs Industry needs

Our analysis of global industry reports reflected growing awareness and demand for ES both in terms of organizational factors and EPIs (see Table 3). Due to the relevance of various BPM concepts, we identified numerous research opportunities at the intersection of BPM and ES, some of which are outlined below.

While the **BPM Lifecycle** has been extended into a Green BPM Lifecycle through several initial studies associated with management and strategy factors of organizations, it is important to have an in-depth understanding of the cultural, strategic and management success factors, and complexities of continuous process improvement towards achieving ES with Green BPM lifecycle phases. Thus, future BPM studies in the context of ES might focus on a suitable lifecycle for designing, evaluating, implementing and monitoring environmentally sustainable processes that meet the environmental sustainability strategies for all EPIs.

While **BPM Architecture Extensions** have not been considered in the context of any EPIs, there is a need to consider further dimensions to current BPM architecture that address EPIs and consolidate processes, practices and structure of organizations into a capability model. Moreover, developing and refining the **Green Capability Maturity Model** will enable organizations to identify, assess and improve their practices, processes and strategies with regards to their ES objectives. There is a scarcity of research that studies suitable Capability Maturity Models that address EPIs.

**Process Performance Measurement** has been applied more than any other BPM concept. There are several studies focusing on measuring and managing EPIs such as Energy consumption, $CO_2$ and GHG emission. However, there is a need to also study the applicability, change and impact on organizational performance after implementation of such techniques and measurement methods. Considering the industry call for action to stop climate change and reduce emissions, studies on implementing measurement methods and required planning, organizational procedures and change in organizational culture are required in the future in the context of all EPIs.

Many researchers have suggested **Process Modeling Notations** to reflect EPIs such as $CO_2$ and GHG in process models. However, further study is required to determine whether these notations are useful in practice and whether they also apply to other EPIs.

Application of **Business Process Reengineering, Design and Optimization** requires further study in terms of most applicable approaches and analysis of implementation success. Stolze et al. (2012) and Thies et al. (2012) provided an initial discussion of potential success factors of implementing environmental sustainable processes. However, given that designing, adopting and implementing environmental sustainability concepts in organizations is associated with costs and significant change(Adger et al. 2005; Christmann 2000), it is critical that future research identify and empirically validate success factors of Green BPM. In particular, we see a lack of guidance and studies on adoption of green practices/processes without compromising organizational and process competence.

In general, we also see a lack of studies on cultural factors at the intersection of BPM and ES (see Table 4). Adoption of ES practices requires a culture shift across organizations (Harris and Crane 2002) and culture change can be achieved through a continuous improvement approach (Gao and Low 2014). Thus, benefits can be achieved from applying BPM initiatives due to its incremental improvement approach and holistic view of people and systems (Pritchard and Armistead 1999; vom Brocke and Sinnl 2011).

## 5   Conclusion

In this paper we aimed to understand the extent to which BPM research has contributed to Environmental Sustainability. To this end, we conducted a literature review, which identified, collected and analysed 36 relevant articles from academic research. Through the analysis we identified the BPM concepts that were applied in the ES context and studied the main ES concepts that were addressed by BPM research. By reviewing authoritative global and industry ES reports, we were able to establish the extent to which BPM academic research has responded to industry and global demand for new ES practices. The majority of studies considered are conceptual contributions in initial stages of development with only a limited number of studies having validated their approaches, thus requiring further exploration. In general, it appears that Green BPM research is in its infancy. In addition to the ES concepts that have received some attention from BPM researchers, we identify further research areas that are of interest to industry but have received no attention from the BPM academic community (see Table 4) as well as several avenues for future research.

This paper is not without limitations. While the identification of the papers was conducted by one researcher and checked by another, one researcher conducted the coding of the papers to BPM and ES concepts. Despite the coding being conducted through multiple iterations of full readings of relevant papers, having a single coder remains a shortcoming that might introduce bias in the analysis. In addition, while we strived to consider the largest set of relevant papers possible, our focus was on contributions published in English only. Our search also resulted in a number of green Supply Chain Management (SCM) publications. We understand that "process" is a studied term in SCM, yet none of the articles found had explicit focus on green BPM or green SCM, therefore green SCM was excluded from this study. Eliminating green IS and IT to focus on green BPM is another limitation, which we are working on by extending our analysis.